\documentclass[preprint,fleqn,showpacs,showkeys]{revtex4}
\usepackage{graphicx}
\usepackage{amssymb}
\usepackage{amsmath}
\usepackage{bm}

\begin{document}
\setcounter{page}{1}

\title[]{The Current-Phase Relation of Ferromagnetic Josephson Junction Between Triplet Superconductors}

\author{Chi-Hoon \surname{Choi}}

\affiliation{Department of Nanophysics, Gachon University, Seongnam 13120, Korea}

\begin{abstract}
We study the Josephson effect of a $\rm{T_1 F T_2}$ junction, consisting of spin-triplet superconductors (T), a weak ferromagnetic metal (F), and ferromagnetic insulating interfaces.
Two types of the triplet order parameters are considered; $(k_x +ik_y)\hat{z}$ and $k_x \hat{x}+k_y\hat{y}$.  
We compute the current density in the ballistic limit by using 
the generalized quasiclassical formalism developed to take into account the interference effect of the multilayered ferromagnetic junction.
We discuss in detail how the current-phase relation is affected by orientations of the d-vectors of superconductor and the magnetizations of the ferromagnetic tunneling barrier. 
General condition for the anomalous Josephson effect is also derived.
\end{abstract}


\maketitle

\section{INTRODUCTION}

The interplay between ferromagnet and spin-singlet superconductor leads to many interesting phenomena in the Josephson effect, including the long-range spin-polarized supercurrent and a reversal of the sign of supercurrent.\cite{r1a,r1b} 
In particular, inhomogeneous distribution of the magnetizations in the ferromagnetic tunneling barrier have a drastic effect on the current-phase relation (CPR):
(i) The second harmonic can become dominant due to the coherent transport of two Cooper pairs.
(ii) The anomalous Josephson effect (AJE): The supercurrent can flow at zero phase difference. (iii) The $\phi$-junction: The ground state of the junction can occur at an arbitrary phase difference.
These features can play an important role in the applications of the quantum electronic devices.\cite{r2a,r2b,r2c}

The Josephson junctions of the spin-triplet superconductor have attracted much attention recently since the discovery of the triplet superconductivity in several materials such as $\rm{Sr_2 RuO_4}$ and the heavy fermion superconductors.\cite{r4a} 
The various types of junctions combining ferromagnet and the triplet superconductors with different pairing symmetries have been studied.\cite{r5a,r5b,r6a,r6b}
The current-phase relation is determined largely by relative orientation between the magnetization of the ferromagnet and the d-vector of triplet superconductor. 
Depending on its orientation, the leading order harmonic in the spectral decomposition of the CPR can be $\sin\phi$, $\cos\phi$, or $\sin 2\phi$, where $\phi$ is the phase difference between two superconductors. 
In the presence of the ferromagnetic barrier, the spin-flip scattering of a magnetic moment can rotate the d-vector, or induce a spin-singlet pairing amplitude, leading to the changes of the CPR.\cite{r5a}
 
In this paper, we consider the triplet superconductor junctions having a multilayered ferromagnetic tunneling barrier.
In the most of the previous works, the barrier has been treated as a uniform ferromagnetic layer.
The schematic diagram of the $\rm{T_1 F T_2}$ junction is depicted in the insert of Fig. 1.
A ferromagnetic metal layer (F) and two ferromagnetic insulating interfaces (I) are surrounded by the two half-infinite triplet superconductors $\rm{T_1}$ and $\rm{T_2}$.
The interfaces are modeled by a delta-function like potential which can incorporate both charge and magnetic scatterings of quasiparticles.  
For the triplet superconductors, we choose the following two typical types of the p-wave order parameters considered in $\rm{Sr_2 RuO_4}$:\cite{r4a}
\begin{eqnarray}
\rm{T_A}: &\;& \mathbf{d}(\mathbf{k})=\Delta_0 (k_x +ik_y )\hat {z}, 
\nonumber \\
\rm{T_B}: &\;& \mathbf{d}(\mathbf{k})=\Delta_0 (k_x\hat{x} + k_y\hat{y}).
\end{eqnarray}
The order parameters have an isotropic gap on the cylindrical Fermi surface.

We compute the current density in the ballistic limit for two kinds of triplet junctions of $\rm{T_A F T_A}$ and $\rm{T_A F T_B}$ while varying the junction parameters such as the magnetization of the interfaces, the exchange field of the ferromagnetic metal layer, and the interlayer thickness. 
We utilize the general formalism of A. Millis \textit{et al.} to treat properly the interference effect from scattering of quasiparticles at the multilayered spin-active interfaces.\cite{r7a}
We investigate in detail how inhomogeneous distribution of the magnetization in the tunneling barrier affects the key features of the ferromagnet-superconductor junction such as the $0-\pi$ transition, the $\phi$-junction, and the anomalous Josephson current.

Before going into details, we summarize our main results.
(i) For the $\rm{T_A F T_A}$ junction, the AJE occurs when the d-vector of $\rm{T_A}$ has both parallel and perpendicular components to a plane formed by the barrier magnetizations. 
(ii) For the $\rm{T_A F T_B}$ junction, the AJE occurs when the barrier magnetizations have a perpendicular component  to the plane spanned by the d-vectors of $\rm{T_A}$ and $\rm{T_B}$.
The $\sin\phi$-term in the CPR can appear when the barrier magnetizations have components along both d-vectors.
(iii) The analytical expression for the effect of spin-flip scattering on the d-vectors is derived in Eq. (14), which can provide a basis to understand the changes of CPR by the barrier magnetization.

\begin{figure}
\includegraphics[width=10cm]{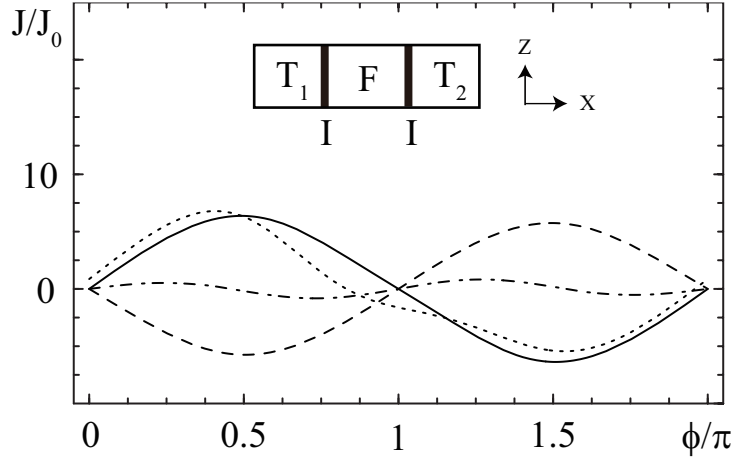}
\caption{Plots of the current density as a function of 
the phase difference $\phi$ for the $\rm{T_A F T_A}$ junction.
The junction parameter, defined by Eq. (10), is chosen as $\bar{U}_{J}=(1,\:\hat{z},\: 1)$, $(\hat{x},\:\hat{z},\: \hat{y})$, $(\hat{z},\:0,\: \hat{z})$, and $(\hat{z},\:\hat{x},\: \hat{z})$ for the solid, dotted, dashed, and dash-dotted curves, respectively. 
The thickness of the ferromagnetic layer is $\bar{d}=0.5$. 
In the insert, the $\rm{T_1 F T_2}$-type junction with a ferromagnetic metal layer surrounded by the two triplet superconductors $\rm{T_1}$ and $\rm{T_2}$ is drawn.} 
\end{figure} 

\section{FORMALISM}

To do our calculation, we follow closely the formalism and the notations of Ref. 11.
First, we summarize the main results of the formalism. 
We consider the superconductor-ferromagnet junction with the specular interfaces located at the positions of $x=0$ and $x=d$, as in Fig. 1.  
The Keldysh Green's function $\hat{G}(x,x')$ 
is decomposed into the four possible combinations of incoming and outgoing waves:  
\begin{equation}
\hat{\tau}_{3}\hat{G}^S(x,x')=\frac{1}{v^{S}}\underset{\alpha,\beta=\pm1}
{\sum}\hat{C}^{S}_{\alpha\beta}(x,x')e^{ip^{S}(\alpha x-\beta x')},
\end{equation}
where the superscript $S$ indicates the layers of the bulk superconductors or the ferromagnet, and the subscripts $\alpha$ and $\beta$ are the indices representing the direction of momentum along the x-axis. 
The magnitude of the Fermi momentum along the x-axis is denoted by
$p^{S}=[2mE_{f}^{S}-(p_{\parallel}^{S})^{2}]^{1/2}$, 
and its corresponding Fermi velocity is $v^{S}=p^{S}/m$.

The differential equations for the envelope functions of one variable, 
$\hat{C}_{\alpha\beta}^{S}(x)=\hat{C}_{\alpha\beta}^{S}(x,x+0)$, 
are derived from the Gorkov equation:
\begin{align}
& \{ i\varepsilon_{n}\hat{\tau}_{3} 
 -\hat{H}^{S} (x, \alpha p^{S})\} 
\hat{C}^{S}_{\alpha \beta}
\nonumber \\ 
& \mp \hat{C}^{S}_{\alpha \beta}
\{ i\varepsilon_{n}\hat{\tau}_{3}-\hat{H}^{S} (x, \pm \alpha p^{S})\}
+ i\alpha v^{S}\partial_{x}\hat{C}^{S}_{\alpha\beta} =0,
\end{align}
where $\varepsilon_{n}=\pi T(2n+1)$ is the Matsubara frequency, and $\hat{\tau}_{i}$ and $\hat{\sigma}_{i}$ are the Pauli matrices in the particle-hole and the spin spaces, respectively.
The upper (lower) sign is for $\alpha=\beta$ ($\alpha \neq \beta$).
The Hamiltonian includes the order parameter $\hat{\Delta}^S $ and the external potential $\hat{V}^S$: $\hat{H}^S = \hat{\Delta}^S+ \hat{V}^S$.
The boundary condition for the interface, which supplements the equations for $\hat{C}_{\alpha\beta}$, can be written by using the transfer matrix $\hat{M}$: 
\begin{equation}
\hat{C}_{\alpha \beta}^L =\sum_{\mu, \nu} \hat{M}_{\alpha \mu} 
\hat{C}_{\mu \nu}^R \hat{M}^{\dagger}_{\nu \beta},
\end{equation} 
where the superscripts L and R denote the left- and the right-hand sides of the interface.
 
Now, we apply the formalism to our $\rm{T_1 F T_2}$-type junction. 
The superconducting order parameter can be written as\cite{r7b}
\begin{equation}
\hat{\Delta}=\left(\begin{array}{cc}
0 & [d_0(\hat{\mathbf{p}})+\mathbf{d}(\hat{\mathbf{p}})\cdot\boldsymbol{\sigma}]i\sigma_{y}\\
i\sigma_{y}[d_0^{*}(-\hat{\mathbf{p}})-\mathbf{d}^{*}(-\hat{\mathbf{p}})\cdot\boldsymbol{\sigma}] & 0
\end{array}\right).
\end{equation}
The d-vectors of triplet superconductor are given by Eq. (1).
The ferromagnetic metal is modeled by a weak exchange field 
$\mathbf{h}$:
\begin{equation}
\hat {V} = \mathbf{h} \cdot \hat{\mathbf{S}} =\left( \begin{array}{cc}
\mathbf{h} \cdot \boldsymbol{\sigma} & 0 \\
0 & -\sigma_{y} \mathbf{h} \cdot \boldsymbol{\sigma} \sigma_y
\end{array}\right). 
\end{equation}
The transfer matrix for the interface can be derived from the scattering theory:\cite{r7c}
\begin{equation}
\hat{M} = \left( \begin{array}{cccc}
\hat{1}+i\hat{v} & 0 & i\hat{v} & 0 \\
0 & \hat{1}+i\hat{v}^* & 0 & i\hat{v}^*\\
-i\hat{v} & 0 & \hat{1}-i\hat{v} & 0 \\
0 &-i\hat{v}^*& 0 & \hat{1}-i\hat{v}^*   
\end{array} \right),
\end{equation}
where $\hat{v} = v_{0} \hat{1}+\mathbf{v}_{m}\cdot \boldsymbol{\sigma}$.
The charge and the magnetic scattering potentials of the interface are denoted by $v_{0}$ and $\mathbf{v}_{m}$, respectively.
We assume that the Fermi velocity $v_f$ is the same everywhere and the magnetic potential is proportional to the magnetization $\mathbf{m}$ of the ferromagnetic interface.

After solving Eq. (3) for $\hat{C}_{\alpha\beta}$ with the proper boundary conditions for the interfaces at x=0 and d, one can calculate the current density. For our translationally-invariant planar interfaces, the current flows along the x-axis.
The current density from the particles incident with the momentum $\mathbf{p}$ can be obtained by\cite{r7a}
\begin{equation}
J(\hat{\mathbf{p}})=\frac{\pi}{2}N_{f}v_{f}
T\underset{n \geq 0}{\sum} (\hat{\mathbf{x}}\cdot\hat{\mathbf{p}}) \mathrm{Tr}
[\hat{\tau_{3}}(\hat{C}_{++} - \hat{C}_{- -} )], 
\end{equation}
where $N_{f}$ is the density of states at the Fermi energy.
The total current density can be computed by integrating the current density $J(\hat{\mathbf{p}})$ over the Fermi surface with $\hat{\mathbf{p}} \cdot \hat{\mathbf{x}} >0$.
The current is continuous at the interface due to particle conservation. 
The free energy of the junction per unit area of the cross section  can be calculated from the current density: 
\begin{equation}
E(\phi)=\frac{\Phi_{0}}{2\pi}\int_{0}^{\phi}J(\chi)d\chi,
\end{equation}
where $\Phi_{0}$ is the flux quantum.\cite{r1a} 

\section{RESULTS AND DISCUSSION}

We now present our numerical calculations. 
We compute the current density for the triplet superconductor junctions while varying the magnetization of the ferromagnetic tunneling barrier.
The superconductors are assumed to have the same uniform gap $\Delta_{0}$.
The energy and the length are scaled in units
of the superconducting gap $\Delta_{0}$ and the superconducting coherence length
$\xi=\hbar v_{f}/\Delta_{0}$, respectively.
To simplify our notations, we introduce a junction parameter  $U_{J}$ for the interface potentials and the exchange field:  
\begin{equation}
U_{J}=(U_L,\:\mathbf{h},\: U_R)=(v_{0}^L+\mathbf{v}_{m}^L,\:  \mathbf{h},\: v_{0}^R+\mathbf{v}_{m}^R), 
\end{equation} 
where L and R denote the left- and the right-hand side interfaces, respectively.
We define the following dimensionless quantities: 
the interlayer thickness $\bar{d}=d/\xi$,
the Fermi wave vector $\bar{k}_{f}=k_{f}\xi$,
the interface potential $\bar{U}_{S}=U_{S}/(\hbar v_{f})$,
the exchange field $\bar{\mathbf{h}}=\mathbf{h}/\Delta_{0}$,
and the junction parameter $\bar{U}_J$.
In our calculation, we set $\bar{k}_{f}=1000$ and the temperature $T=0.1 \Delta_{0}$.

We compute the current density as a function of the phase difference between two superconductors in the ballistic limit by using Eq. (8) for a normal incidence of particles whose momentum is parallel to the x-axis.
The current density is normalized by $J_{0}=N_{f}v_{f} T/4$. 
For the normal incidence, the d-vectors for the superconductors $\rm{T_A}$ and $\rm{T_B}$ are aligned to the z- and the x-axes, respectively.
For a different angle of incidence, it is straightforward to generalize the calculation by replacing the position variable $x$ with $x/\cos\theta_k$, with the angle $\theta_k$ being measured from the normal to the interface. 

First, we discuss briefly the effect of the magnetic scattering potential on the pairing amplitude to understand its effect on the CPR.\cite{r5a}
We assume that particles are incident from left of the interface. 
The Green's function for the transmitted particles to the right of the interface can be obtained from the boundary condition of Eq. (4):
\begin{equation}
\hat{C}_{++}^R=\hat{K}_{++} \hat{C}_{++}^L 
\hat{K}_{++}^{\dagger},
\end{equation}
where $\hat{K}_{++}=\hat{M}_{++}^{-1}$. 
The pairing amplitude $\hat{f}^R$ of the transmitted Green's function $\hat{C}_{++}^R$ is related to the pairing amplitude $\hat{f}$ of the incident one $\hat{C}_{++}^L$: 
\begin{equation}
\hat{f}^R=\hat{K} \hat{f}\hat{K}^{*},
\end{equation}
where $\hat{f} = f_0 + \mathbf{f} \cdot \boldsymbol{\sigma}$ and 
\begin{equation}
\hat{K}=\frac{(1+iv_0)\hat{1} - i\mathbf{v}_{m} \cdot \boldsymbol{\sigma}}{(1+iv_0)^2+v_m^2}.
\end{equation} 
Up to the first order in $v_m$, the transmitted order parameter $\hat {f}^R$ can be expanded as
\begin{align}
f_0^R&=\gamma f_0 -2i\gamma^2\mathbf{v}_{m} \cdot \mathbf{f},
\nonumber \\
\mathbf{f}^R&=\gamma \mathbf{f} -2 i \gamma^2 [f_0 \mathbf{v}_{m} + v_0 \mathbf{v}_m \times \mathbf{f}],
\end{align}
where $\gamma=(1+v_0^2)^{-1}$.
The nonmagnetic scattering potential $v_0$ just affects the magnitude of the incident order parameter by a factor of $\gamma$.
The magnetic potential $\mathbf{v}_{m}$ can induce the singlet component $f_0^R$ via the term $\mathbf{v}_{m} \cdot \mathbf{f}$ and the triplet component $\mathbf{f}^R$ via the terms of $f_0 \mathbf{v}_{m}$ and $v_0 \mathbf{v}_{m} \times \mathbf{f}$.
The induced pairing amplitudes acquire a $90^0$ phase shift from the spin-flip scattering.
We remark that the exchange field of the ferromagnetic metal layer has the similar effect on the d-vector as the magnetic scattering potential of the interface.

In Fig. 1, we calculate the current density as a function of the phase difference for the $\rm{T_A F T_A}$ junction. 
When the junction barrier is nonmagnetic or weakly magnetic, as in the solid curve, a Cooper pair of the bulk superconductor can directly tunnel through the barrier and the current is dominated by the $\sin\phi$-term in the spectral decomposition of the CPR.
As the magnetic scattering becomes larger, as in the dashed curve, the current can reverse its sign as in the case of the spin-singlet superconductor junction.\cite{r1a,r1b}
When the exchange field is introduced in the direction perpendicular to the d-vector, as in the dash-dotted curve, the second harmonic term can be larger than the first harmonic because the exchange field in the ferromagnetic layer can block the tunneling of a Cooper pair.

When the direction of magnetization at the two interfaces and the exchange field of the ferromagnetic layer are mutually perpendicular to each other while the exchange field is aligned to the d-vector, as in the dotted curve, the $\cos\phi$-term appears in the CPR, leading to the anomalous Josephson current.
In general, the AJE occurs for a non-coplanar distribution of the barrier magnetizations. When the barrier magnetizations have a coplanar distribution, the AJE occurs if the d-vector of the triplet superconductor has  both parallel and perpendicular components to the plane formed by the barrier magnetizations. 
In our choice of the barrier magnetization $\bar{U}_J=(\hat{x},\: \hat{z},\: \hat{y})$, a $f_x$-component of the pairing amplitude can be induced in the ferromagnetic layer by the magnetization $m_y$ of the right interface and also by the successive spin-flip scatterings from the magnetization $m_x$ of the left interface and the exchange field $h_z$. A tunneling between the induced $f_x$-components, whose phases differ by  $90^0$, leads to the $\cos\phi$-harmonic.

\begin{figure}
\includegraphics[width=10cm]{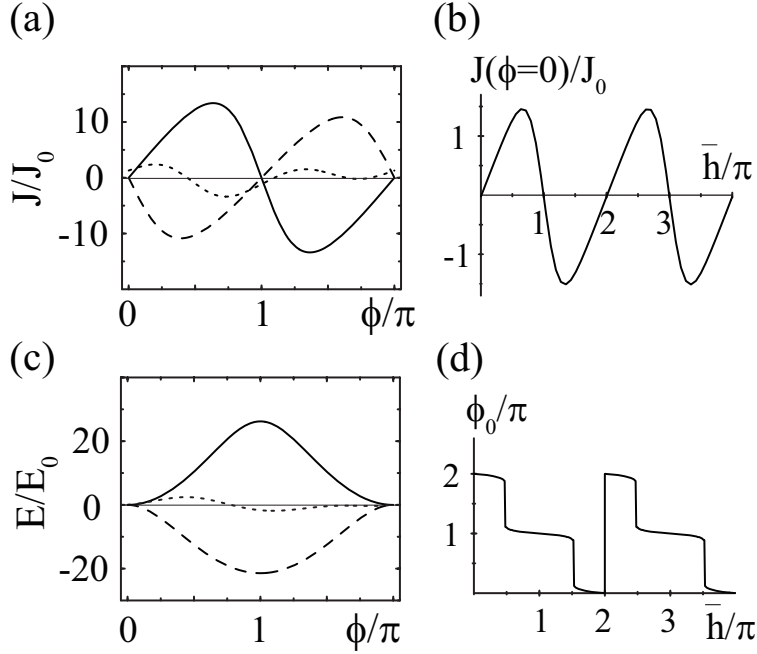}
\caption{Plots of (a) the current density and (c) the free energy of the junction as a function of the phase difference $\phi$ for the $\rm{T_AFT_A}$ junction. The junction parameter is chosen as $\bar{U}_{J}=(\hat{x} ,\:\bar{h} \hat{z},\: \hat{y})$ and the interlayer thickness is $\bar{d}=0.5$. The magnitude of the exchange field $\bar{h}$ is 0, $0.5\pi$, and $\pi$ for the solid, dotted, and dashed curves, respectively.
In (b) and (d), the anomalous Josephson current density $J(\phi=0)$ and the ground state phase $\phi_0$ are plotted as a function of $\bar{h}$. }
\end{figure} 

In Fig.2, we compute the current density and the free energy of the junction as a function of the phase difference for the $\rm{T_A F T_A}$ junction to study the effect of the exchange field on the CPR. 
The exchange field is chosen to be parallel to the z-axis and the magnetizations of the left and the right interfaces are parallel to the x- and the y-axes, respectively.
In the absence of the exchange field, the middle layer becomes a normal metal and the dominant harmonic is $\sin\phi$.
For a finite exchange field, the junction can have the $\cos\phi$-term as discussed in Fig. 1, leading to the AJE.
In Fig. 2(b), we plot the anomalous Josephson current as a function of the magnitude of the exchange field. 
It shows an oscillation of the period of $2\pi$ due to a resonant scattering of quasiparticles in the ferromagnetic layer between the two surrounding interfaces.  
In the ballistic limit, quasiparticles gain a phase factor of $e^{2 i hd/(\hbar v_f)}$ in the ferromagnetic layer during the process of a Cooper pair tunneling.\cite{r2a}

In Fig. 2(c), we calculate the free energy of the junction by using Eq. (9).
The unit of the free energy is given by $E_0=\Phi_0 / 2\pi$.
The minimum of the junction energy occurs at $\phi_0$ = 0, $1.1\pi$, and $\pi$ for $\bar{h}$ = 0, $0.5\pi$, and $\pi$, respectively.
In Fig. 2(d), the ground state phase $\phi_0$ is plotted as a function of the magnitude of the exchange field, which shows the oscillation of the period of $2\pi$, the $0-\pi$ transition, and the $\phi$-junction.

\begin{figure}
\includegraphics[width=10cm]{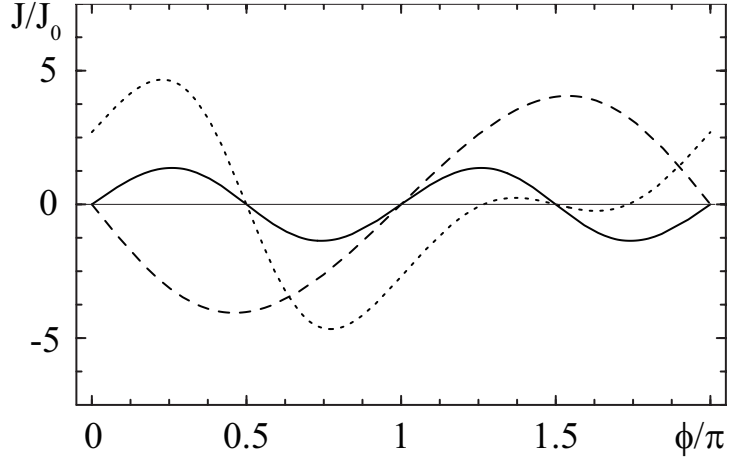}
\caption{Plots of the current density as a function of 
the phase difference $\phi$ for the $\rm{T_A F T_B}$ junction.
The junction parameter is chosen as $\bar{U}_{J}=(1,\:\hat{z},\: 1)$, $(\hat{y},\:0,\: \hat{y})$, and $(1.5\hat{z},\:0,\: 1.5\hat{x})$ for the solid, dotted, dashed curves, respectively. 
The thickness of the ferromagnetic layer is $\bar{d}=0.5$.}
\end{figure} 

In Fig. 3, we plot the several distinctive types of the CPR for the $\rm{T_A F T_B}$ junction.
The d-vectors of the superconductors $\rm{T_A}$ and $\rm{T_B}$ are aligned to the z- and the x-axes, respectively.
When the tunneling barrier is nonmagnetic, the second harmonic $\sin 2\phi$ is the leading order in the CPR due to the orthogonality of the d-vectors.
When the exchange field is applied in the x-z plane, the second harmonic remains to be dominant, because the induced pairing amplitude by the exchange field is orthogonal to those of the bulk superconductor.
For example, the exchange field parallel to the z-axis, as in the solid curve, can induce the pairing amplitudes $f_0$ and $f_y$ which are orthogonal to the d-vectors of the bulk superconductor. 
In the dashed curve of Fig. 3, we choose the magnetization at each interface to be parallel to the d-vector of its adjacent superconductor.
Both interfaces can induce the same singlet component $f_0$ in the ferromagnetic layer.
A Cooper pair tunneling between the induced singlet components leads to the $\sin\phi$-harmonic.
The $\sin\phi$-harmonic can generally appear when the interface magnetizations or the exchange field have components along both d-vectors of $\rm{T_A}$ and $\rm{T_B}$.

The AJE can occur when the magnetization at the two interfaces or the exchange field have a perpendicular component  to the plane spanned by the d-vectors of $\rm{T_A}$ and $\rm{T_B}$.
In our choice of $\bar{U}_J=(\hat{y},\: 0,\: \hat{y})$, the magnetization along the y-axis can induce a pairing amplitude $f_x$ from the superconductor $\rm{T_A}$.
A Cooper pair tunneling between the induced pairing amplitude and the superconductor $\rm{T_B}$, whose phases differ by $90^0$, leads to the $\cos\phi$-harmonic.
We remark that the magnetization at the one interface alone is not enough for the AJE unless there is another nonvanishing element in the interface potentials.

\begin{figure}
\includegraphics[width=10cm]{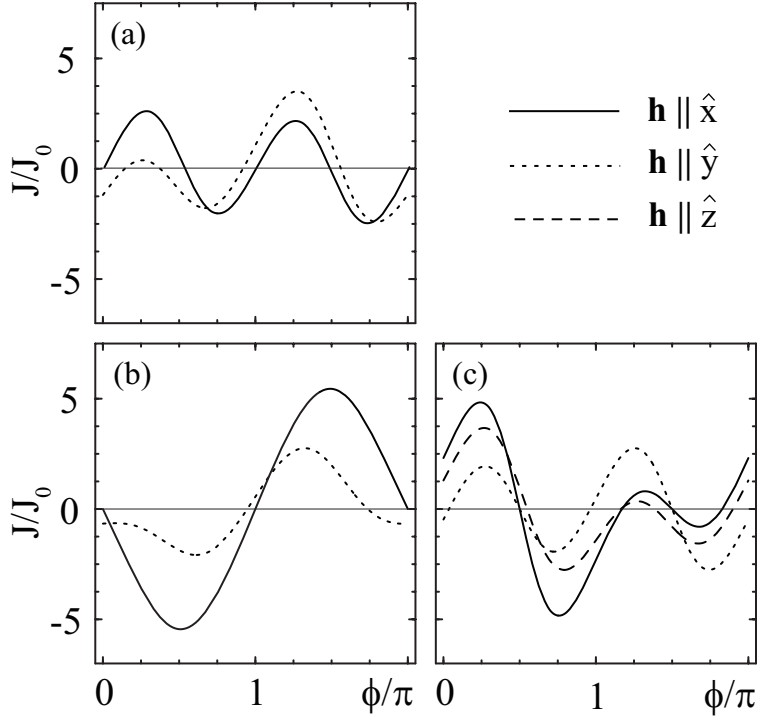}
\caption{Plots of the current density as a function of 
the phase difference $\phi$ for the $\rm{T_A F T_B}$ junction.
Three different exchange fields are chosen; $\bar{\mathbf{h}}$ = $\hat{x}$, $\hat{y}$ and $\hat{z}$.
The interface magnetizations are $(\bar{U}_{L},\:\bar{U}_{R})$ =
$(\hat{x},\:\hat{z})$, $(\hat{z},\:\hat{x})$, 
and $(\hat{x},\:\hat{y})$ for (a), (b), and (c), respectively.
In (a) and (b), the current densities for the exchange field along the x- and the z-axes are identical.}
\end{figure} 

In Fig. 4, we calculate the current density of the $\rm{T_A F T_B}$ junction for several different sets of orientations of the interface magnetization and the exchange field.
For the interface magnetization of $(\bar{U}_L,\:\bar{U}_R )=(\hat{x},\:\hat{z})$, as in Fig. 4(a), the current is dominated by the second harmonic $\sin 2\phi$ over the first harmonic $\sin\phi$.
When the magnetizations for the right and left interfaces are interchanged as $(\bar{U}_L,\:\bar{U}_R )=(\hat{z},\:\hat{x})$, as in Fig. 4(b), the magnetization at both interfaces that are next to the superconductors can induce a fairly large $f_0$-component simultaneously.
A Cooper pair tunneling through the $f_0$-component can make the $\sin\phi$-term become larger than the second harmonic term.
In Fig. 4 (c), the interface magnetization is $(\bar{U}_L,\:\bar{U}_R )=(\hat{x},\:\hat{y})$.
The anomalous Josephson current appears independent of the orientations of the exchange field due to the magnetization along the y-axis.
For the exchange field along the x- and the y-axes, the current depends mainly on the harmonics of $\cos\phi$ and $\sin 2\phi$, while the $\sin\phi$-term appears along the z-axis.

In conclusion, we study effect of the nonuniform distribution of the barrier magnetizations on the current-phase relation for the two types of the triplet superconductor junctions $\rm{T_A F T_A}$ and $\rm{T_A F T_B}$.
The $0-\pi$ transition, the $\phi$-junction, and the AJE can readily be realized by adjusting the relative orientation between the d-vectors of superconductor and the barrier magnetizations.
In general, the condition for the AJE is more restrictive in $\rm{T_A F T_A}$ than in $\rm{T_A F T_B}$.
We also discuss the interference effect on the CPR for the multilayered ferromagnetic junction.
In the future, we plan to extend the present work to calculate other physical quantities such as the induced pairing amplitude and the spin current for the triplet superconductor junctions, while incorporating other types of the unconventional superconductors and the effect of a strong exchange field for the ferromagnet.

\end{document}